**Article**

# Metallization of leaf-derived lignocellulose scaffolds for high-performance flexible electronics and oligodynamic disinfection

Check for updates

Rakesh Rajendran Nair [1] ✉, Mihai Nita-Lazar[2], Valeriu Robert Badescu[2], Cristina Iftode[2], Jakob Wolansky [1], Tobias Antrack[1], Hans Kleemann [1] & Karl Leo[1] ✉

Vascular tubules in natural leaves form quasi-fractal networks that can be metallized. Traditional metallization techniques for these lignocellulose structures are complex, involving metal sputtering, nanoparticle solutions, or multiple chemical pretreatments. Here we present a novel, facile, and reliable method for metallizing leaf-derived lignocellulose scaffolds using silver microparticles. The method achieves properties on-par with the state-of-the-art, such as broadband optical transmittance of over 80%, sheet resistances below 1 $\Omega$/sq., and a current-carrying capacity exceeding 6 A over a 2.5 × 2.5 cm² quasi-fractal electrode. We also demonstrate copper electrodeposition as a cost-effective approach towards fabricating such conductive, biomimetic quasi-fractals. Additionally, we show that these metallized structures can effectively eliminate pathogenic microorganisms like fecal coliforms and E. coli, which are bacterial indicators of microbiological contamination of water. We finally show that these oligodynamic properties can be significantly enhanced with a small externally applied voltage, indicating the noteworthy potential of such structures for water purification and pollution control.

Previous studies have shown that leaf venation can be metallized using sputtering[1] or metal nanoparticle coatings[2–4]. Work from Han et al.[1] was among the first to showcase how metal sputtering on organic quasi-fractal structures like leaf skeletons and spider-webs can result in transparent and conductive micro-scaffold networks. Sharma et al.[5] demonstrated chemical coating of Ag nanowires onto quasi-fractal leaf skeletons, yielding electrodes with 80% transparency and a sheet resistance of 8 $\Omega/\square$, and implemented them as nature-derived flexible, transparent heaters. Furthermore, quasi-fractal structures with conductive properties have previously been shown to demonstrate remarkable efficiency across diverse application scenarios. For instance, electrodes structured in fractal-like shapes have proven to be notably superior for tasks such as neural stimulation and electrophysiological sensing[6–9], outperforming the ubiquitously implemented fully planar electrodes. Additionally, numerous studies have found that mimicking biological quasi-fractal structures in electrode design also leads to enhanced electro-optical performance[10–12] which can directly improve OLED and OPV device efficiencies while reducing the material needed for electrode fabrication.

Here, we demonstrate a novel and facile method to coat leaf venation with silver (Ag) microparticles by pre-treating the lignocellulose fibers using corona discharge treatment (CDT). We further employ environmentally friendly binding agents for the strong adhesion of Ag microparticles to the lignocellulose scaffolds by dispersing the Ag microparticles in the form of a screen-printable ink. The choice to use Ag as the metallizing agent not only stems from Ag having high conductivity (and its ability to retain this conductivity post oxidation in air), but also due to the oligodynamic nature of Ag metal and its ubiquitous use in the preparation of conducting inks for printing. The final coated lignocellulose structures result in mechanically robust free-standing quasi-fractal electrodes with sheet resistances consistently below 1 $\Omega/\square$ and optical transmittance values of 80%. Figure 1a shows the image of a fresh mature leaf of the Magnolia tree (Magnolia liliiflora). The mesophyll of the leaf is removed to expose the quasi-fractal venation of the xylem and phloem tubules (Fig. 1b). Once this microstructure is coated with a thin layer of Ag micro-particles (details in Methods), a highly conducting, quasi-transparent, and flexible electrode is achieved, as shown in Fig. 1c. An illustration of the Ag microparticles binding to the lignocellulose fibers (naturally found bundled together as macrofibrils) is shown in Fig. 1d.

¹Dresden Integrated Center for Applied Physics and Photonic Materials (IAPP) and Institute for Applied Physics (IAP), Technische Universität Dresden, Dresden, Germany. ²National Research and Development Institute for Industrial Ecology-ECOIND, Bucharest, Romania. ✉e-mail: rakesh_rajendran.nair@tu-dresden.de; karl.leo@tu-dresden.de

**Fig. 1 | Ag coating of leaf-derived lignocellulose scaffolds. a** Pristine Magnolia leaf sample (**b**) Leaf after processing to expose lignocellulose structure (**c**) Lignocellulose structure coated with Ag microparticles dispersed in binder (**d**) Illustration of the metallized lignocellulose macrofibrils.

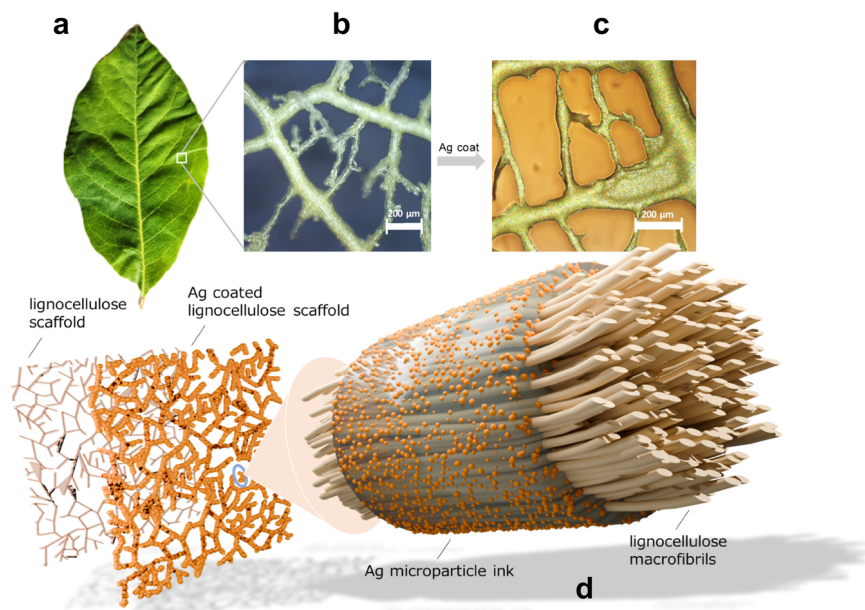

It is known that silver metal (among some other heavy metals), across its various forms, exerts a biocidal effect[13,14] which is termed as the oligodynamic effect. The oligodynamic properties of Ag have yet to be fully harnessed in the field of flexible electronics and represent a promising avenue for addressing issues related to microbial contamination of not just surfaces but also stored drinking water. Coliform bacteria are gram-negative microbes that are prevalent across diverse environmental substrates, including soil and water, and can potentially pose risks to human health. Their detection in aquatic ecosystems serves as a direct indicator of microbial contamination, serving as an alarm for compromised water quality, particularly in areas designated for water supply. A facile, low carbon-footprint technique to remove such pathogens without introducing further, potentially toxic chemicals into the environment is a valuable avenue of research.

Therefore, in this work, we first demonstrate a novel method to coat Ag on to leaf-derived natural lignocellulose quasi-fractals and compare the properties of the resulting freestanding, flexible conducting scaffolds with respect to values obtained using previously reported coating techniques. The first half of this work demonstrates multiple characterization techniques to show that the method of metallization reported here results in values on par with the state-of-the-art of leaf-skeleton based conducting quasi-fractal fabrication. Our use of microparticles additionally allows this technique to evade the ecotoxicity risks associated with the use of Ag nanoparticles[15–17]. Previous reports have largely focussed on implementing these conducting quasi-fractal structures as flexible heaters or quasi-transparent electrodes with impressive results[1–5], therefore we additionally perform conductivity and transparency measurements to show that the technique reported here is also perfectly suitable for such implementations.

Having used the first half to conclusively demonstrate the effectiveness of the metallization technique developed in this work, we push the state-of-the-art further by showing that such bio-derived conducting structures show promise as wastewater treating sieves and may have novel applications in maintaining the purity of stored potable water, especially in arid or drought-sticken areas. For this, we test the oligodynamic efficacy of these Ag coated scaffolds against pathogenic microorganisms with very promising results. We additionally show preliminary findings indicating that the oligodynamic effect might be radically improved by the application of a small voltage across these conducting quasi-fractal structures.

## Methods

Mature Magnolia leaves are freshly procured from local plants and are washed under running water before being placed in an ultrasonic ethanol bath for 10 min to remove any impurities. The leaves were then cut to a size of $2.5 \times 2.5$ cm$^2$ and boiled in DI water for 30 min. The samples are subsequently placed in a 1 M aqueous solution of $Na_2CO_3.10H_2O$ (Sodium Carbonate decahydrate) and heated at a constant temperature of 90 °C under stirring overnight. The samples are then removed and placed into a separate vessel containing water at room temperature which is subsequently placed in an ultrasonic bath for 20 min. The process is repeated until the water no longer changes color after the bathing process. The venation in the leaves is clearly visible at this point and gentle brushing with gloved fingers is enough to remove excess biomass. The leaf-skeletons are subsequently washed with detergent to remove the rest of the mesophyll and then bleached via submergence in 10% $H_2O_2$ for 15 min before being dried and flattened with weights.

### Materials

Branched Polyethylenimine (PEI) and ethyl cellulose (EC) were procured from Sigma Aldrich, Germany. Highly viscous EC polymer solution was prepared by dissolving 8 gm of EC in 36 ml of 2-Butoxyethanol ($C_6H_{14}O_2$) and leaving the mixture under low stirring at 90 °C overnight. Subsequently, 5% (by weight) of rapeseed oil was added to this solution before increasing the temperature to 170 °C for 5 min under stirring for full dissolution before being cooled slowly. This solution was used for the synthesis of Ag ink for screen printing and the required Ag microparticles measuring 2–8 μm were procured from Sigma Aldrich, Germany. $CuSO_4.5H_2O$ (copper sulfate pentahydrate) was procured in powder form and approximately 16 gm was dissolved in 500 ml of DI water at room temperature under constant stirring for the preparation of the electrolysis bath.

### Silver (Ag) coating

In lieu of expensive nanoparticle/nanowire dispersions, the lignocellulose structures are treated with microparticulate Ag ink generally implemented for screen printing purposes. Preparation of the scaffolds is done by first ultrasonicating them for 5 min in a 1:5 mixture of isopropanol and water to remove any impurities and oily residue before drying on a hotplate maintained at 60 °C for 10 min. The scaffolds subsequently undergo corona treatment to induce charged functional groups on the surface of the fibers.

As seen in Fig. 2, the $2.5 \times 2.5$ cm² lignocellulose scaffolds are subsequently dip-coated (using a spatula to apply gentle force) into Ag ink with a viscosity between 45–55 Pa.s. The ink is prepared using 2-Butoxyethanol ($C_6H_{14}O_2$) as a dispersing agent for Ag microparticles with ethyl cellulose as a binder that is dissolved in $C_6H_{14}O_2$ at a mass ratio of 1:5. Immediately after coating, the samples are sandwiched between laboratory tissues and pressure is applied to remove any un-adhered ink. The samples are subsequently dried at 90 °C for 15 min and a flat load is placed on top during the last 5 min of drying.

### Copper (Cu) electroplating

Elemental Cu is deposited using electroplating in a $CuSO_4.5H_2O$ bath. The samples are pre-treated with CDT, and subsequently dipped in Ag ink diluted with 2-Butoxyethanol to a viscosity of 5-10 Pa.s (from 55–65 Pa.s). This results in a very thin coating of Ag on the lignocellulose scaffolds ($R_{sh} > 70 \Omega/\square$). This scant layer of Ag is then used for the electroplating of Cu at 100 mA DC (2 V, current limited mode) against a Copper Clad Laminate (CCL) anode kept 2 cm away from the scaffold, which due to its thin Ag coat, acts as the cathode when placed in a 1 M solution of $CuSO_4.5H_2O$ which functions as the electrolyte.

### Scanning-electron microscopy

SEM images were obtained with a Zeiss Gemini 500 SEM under 1e-5 mbar vacuum and operated at 3 kV. Secondary-electrons were detected by a HE-SE detector placed on the side. To allow the deposited electrons to flow away, the sample was placed on a sticky carbon-pad and on the edge additionally connected with silver paste. For energy-selective backscattering measurements, the ESB grid was set to 1500 V.

### Ag quantification

The silver ion concentration quantification was performed according to the SR EN ISO 17294-2:2017 standard on an Agilent 7900 ICP-MS equipped with an SPS 4 autosampler. ICP multi element standard solution IV for calibration was purchased from Sigma Aldrich and the calibration curves were drawn over the range of 10–50 μg/L. The samples were previously digested with royal water and then subjected to analysis.

### Microbiological analyses

Total coliform and fecal coliform bacteria quantification was performed in duplicate by the Most Probable Number (MPN) method (IDEXX) according to ISO 9308-2 standard [*ISO 9308-2 – Water quality, detection and enumeration of coliform organisms, thermotolerant coliform organisms and presumptive Escherichia coli – Part 2: Multiple tube (most probable number) method*]. Briefly, each sample was homogenized with a Colilert-18 medium, then the mixed solution was dispensed in Idexx bags and incubated at 36 ± 2 °C for 18–22 h for total coliform bacteria and E.coli or at 44.5 ± 0.5 °C for 18–22 h for fecal coliform bacteria. The bacterial quantification was expressed in MPN/100 ml.

## Results and discussion
### Metallization and characterization

Magnolia liliiflora leaves were extracted locally and the mesophyll was removed using an alkaline etching process based on previous report[4] in order to expose the internal quasi-fractal vasculature of the leaf (details in Methods). This vasculature is primarily made up of lignin, cellulose, and hemicellulose, which together are referred to as 'lignocellulose'[18]. In order to coat these fibers with Ag, they are first exposed to corona discharge treatment (CDT) using a 10 KV handheld corona surface treater, which generates a partial anionic charge on the fibers. This, when submerged in an Ag microparticle ink incorporating protonated polyethylenimine (PEI) as an adhesion promoter, allows Ag particles to bind strongly to the lignocellulose microstructures as illustrated in Fig. 3a. CDT is specifically chosen here due to its solvent-free and eco-friendly disposition[19]. The process can additionally be carried out in air without needing special atmospheric conditions while facilitating the express activation of the polymer surface without altering bulk properties. Figure 3a.1 shows the structure of the lignocellulose quasi-fractals. The corona treatment creates activated sites on the lignocellulose surface by deprotonating the hydroxyl (-OH) and carboxyl (-COOH) functional groups, rendering an effective negative charge on the structure as shown in Fig. 3a.2. The fibers when subsequently submerged in Ag ink obtained from 5 different 2.5 × 2.5 cm² electrodes. Samples without the corona treatment begin losing conductivity soon and are completely insulating within 10 min of ultrasonic agitation. However, the corona treated samples retain conduction and the effective resistance only rises by 4Ω even after 18 min in an ultrasonic bath. The data shows that the pre-treatment grants excellent adhesion and thus reliably high conductance to the resulting quasi-fractal electrodes. Additionally, although pristine untreated lignocellulose fibers allow Ag particles to naturally adhere, the results indicate that the adhesion is not stable and fibers loose conductivity easily upon mechanical agitation.

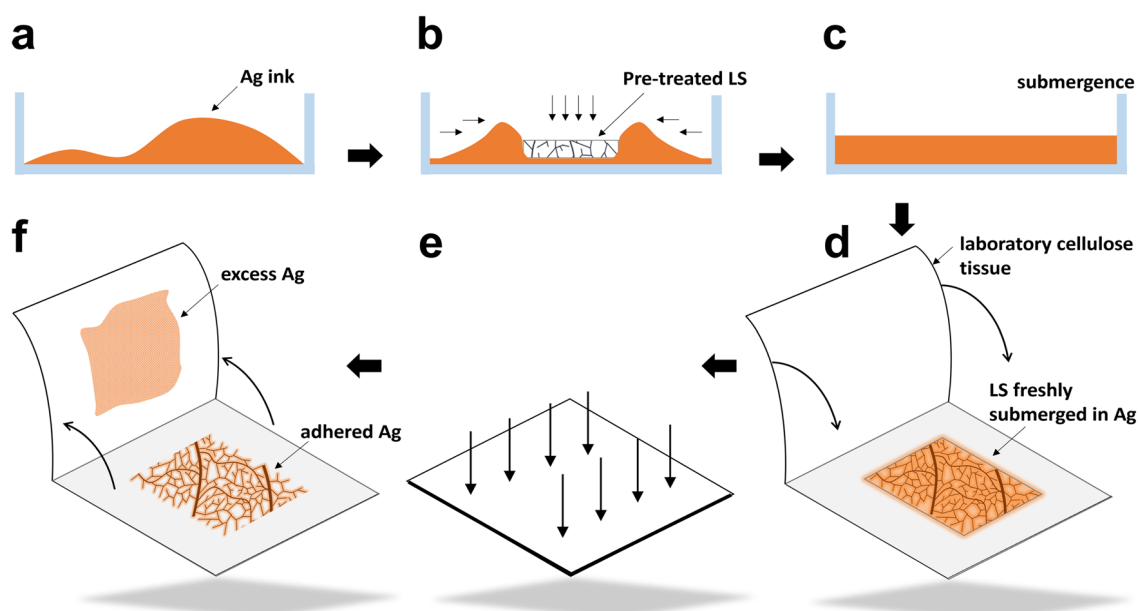

**Fig. 2 | Ag coating process.** (**a**) Ag ink for screen printing (**b**) Leaf skeleton pre-treated with corona discharge treatment (CDT) placed on Ag ink reservoir and pressed in as shown in (**c**) until full submergence. LS is extracted from Ag ink and placed on a laboratory tissue as shown in (**d**) before being sandwiched between two layers of tissue with pressure applied on top as illustrated in (**e**). This causes excess, unadhered Ag to be extracted via absorption into the tissue fibers as shown in (**f**).

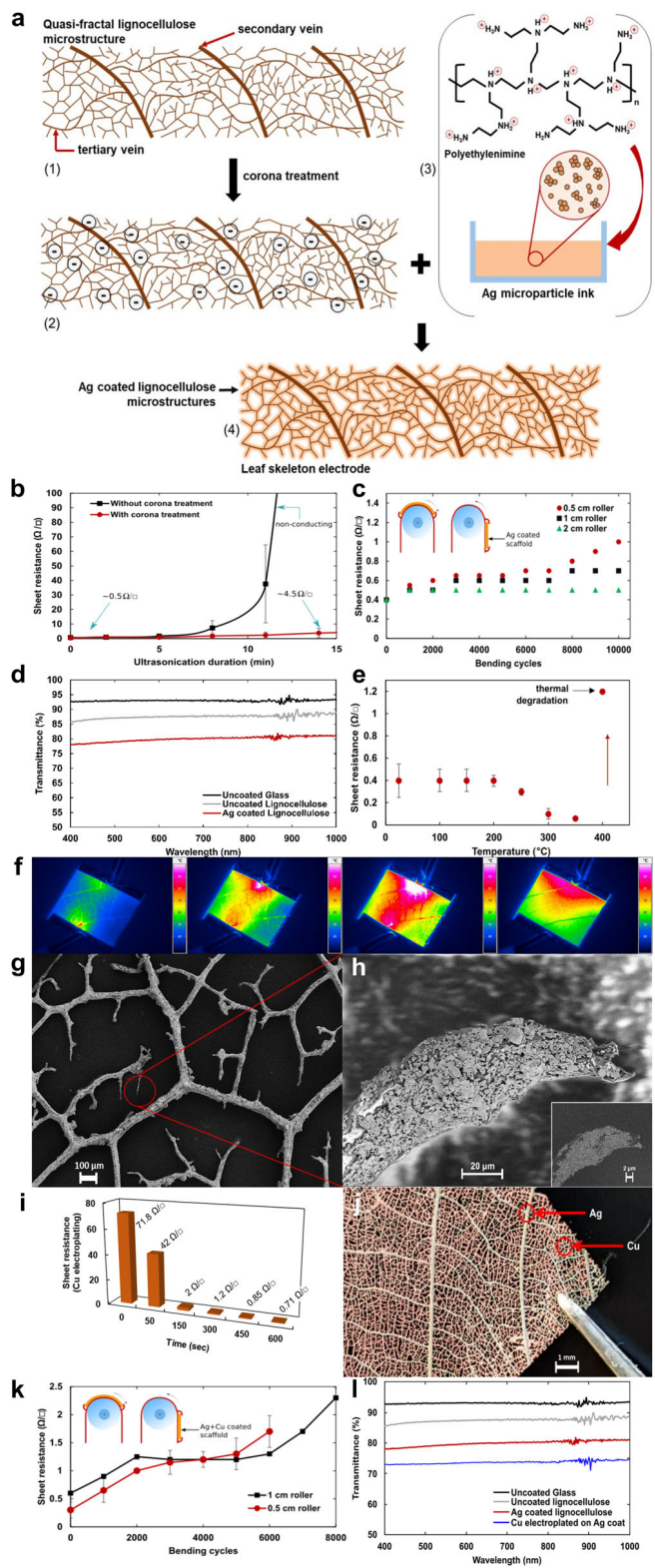

**Fig. 3 | Thermo-electro-mechanical testing of Ag coated lignocellulose structures. a** Illustration of (1) leaf lignocellulose fibers (2) partial negative charge induced onto the lignocellulose fibers after corona treatment (3) protonated PEI in Ag microparticle ink (4) Ag microparticles preferentially adhere to corona treated lignocellulose fibers resulting in a highly conducting, free-standing, quasi-fractal structure. **b** Change in resistance upon ultrasonic agitation of Ag coated lignocellulose structures with and without corona treatment (**c**) Change in sheet resistance during 10,000 bend-relax cycles over 5 mm 10 mm and 20 mm rollers (**d**) Light transmittance measurements of glass, uncoated lignocellulose structures and Ag coated lignocellulose structures (with corona pre-treatment) (**e**) Change in resistance with respect to temperature (**f**) Rise in temperature with respect to 4 A of current flow (1) heat distribution 2 s after application of power (2) 4 s after application of power (3) heat spreading across the entire surface of the electrode 6 s after the application of power (4) the quasi-fractal structure allowing for prompt heat dissipation 5 s after removal of applied power (**g**) SEM image of the Ag coated scaffolds showing their quasi-fractal structure (**h**) Ag particles thoroughly coating the fibers all the way to the edges (scale bar 100 μm) **inset**: the distribution of Ag particles (in gray) using energy selective backscattered (EsB) electron detection (scale bar 2 μm) (**i**) Variation of sheet resistance over the duration of copper electroplating of Ag pre-treated scaffolds (**j**) Image of the Cu coated lignocellulose scaffold (**k**) Change in sheet resistance during 8000 (1 cm roller) and 6000 (0.5 cm roller) bend-relax cycles of the Cu electroplated Ag samples (standard deviation of each data point did not exceed ± 0.5 Ω/sq. (**l**) Transmittance of the leaf scaffolds after Cu electroplating on Ag for 2 min drops from ~80% to ~75%.

Mechanical stability measurements of the Ag coated lignocellulose structures are shown in Fig. 3c and are measured using a custom-built bending machine. The resistance of 2.5 × 2.5 cm² samples is measured at regular intervals while they are bent over metal cylinders with diameters of 5 mm, 10 mm and 20 mm to a total of 10,000 bending cycles for each. It is apparent that the Ag coating after corona treatment results in a robust bonding and a negligible change in sheet-resistance is observed even after 10,000 bending cycles over 10 mm and 20 mm rollers while the resistance only increases from 0.4 Ω to 1 Ω after 10,000 bending cycles over a 5 mm diameter roller.

Light transmittance measurements of the Ag coated lignocellulose structures show a maximum transparency value of 80% at 550 nm (Fig. 3d). Although uncoated glass has the highest transmittance values (ranging round 92%), uncoated, pristine Magnolia leaf-skeletons used in this study have transmittances generally over 85% (Fig. 1b). The gap in transmittance of about 6–7% between the uncoated and Ag-coated structures may be further reduced by optimizing the coating technique to achieve improved transmittance values. In previous studies, the transmittance and sheet resistance of ITO-coated glass has been measured to be about 85%[20] and around 15 Ω/□[21] respectively. With an average sheet resistance of 0.5 Ω/□ and the 80% transmittance of the Ag coated lignocellulose fibers, the figure of merit (FOM) calculation was based on the FOM function developed by Cisneros-Contreras et al.[22], who improved upon Haacke's original figure of merit function[23] for transparent electrodes. The 'FOM Haacke High Resolution' (FOM$_{H\text{-}HR}$) is given as:

$$\phi_{H-HR} = \frac{T}{\sqrt[4]{R_{\square}}}$$ (1)

containing protonated amine groups (Fig. 3a.3) allow the Ag particles to bind to the lignocellulose, resulting in a highly stable and conducting quasi-fractal Ag electrode as shown in Fig. 3a.4.

The adhesion of the Ag particles to the lignocellulose scaffolds with and without the corona treatment is compared by ultrasonicating the samples as shown in Fig. 3b. Here, each data point combines the measurements where '$\phi_{b-HR}$' is the FOM in units of $\Omega^{-1}$ and 'T' is the transmittance at 550 nm. 'R$_\square$' is the measured sheet resistance with 'n' having a recommended value of 10.

The figure of merit for the quasi-fractal lignocellulose electrodes presented in this work is around $0.85\ \Omega^{-1}$. Considering the literature on transparent electrodes, FOMs greater than $0.45\ \Omega^{-1}$ are already deemed suitable for applications in photovoltaics, light emitting diodes (LEDs), gas sensors, thermal collectors etc.[24]. Additionally, the FOM reported in this publication is commensurate with and improves upon some of the state-of-the-art results reported previously using more sophisticated approaches[24–27]. Leaf based electrodes offer further advantages by being fabricated in ambient conditions without the use of energy intensive techniques such as sputtering or physical vapor deposition (PVD). Finally, the quasi-fractal structure of the electrodes also naturally enhances their electro-optical performance as multiple studies into bio-inspired quasi-fractal electrodes have shown[1,10–12].

Next, the thermal response of the Ag coated lignocellulose is studied by heating $2.5 \times 2.5\ cm^2$ samples in open air while monitoring the change in resistance. It can be seen in Fig. 3e that the sheet resistance remains unchanged until about 200 °C after which a drop in resistance occurs due to the sintering of Ag until 300 °C. Any heating beyond this point results in the substrate loosing structural integrity and fracturing at the slightest contact.

Electro-thermal stability of the lignocellulose-Ag electrodes is measured by passing a direct current (DC) from a constant current source up till a value of 4 A while temperature readings are recorded in parallel. Figure 3f shows that the heat generation from 4 A of current flow stays around 100 °C due to the high conductivity of the electrodes along with the excellent heat dissipation provided by the quasi-fractal structure. The temperature was measured simultaneously with a temperature sensor in thermal contact with the electrodes and a non-contact infrared thermometer. Experiments show that the electrodes can effectively handle currents as high as 6 A indefinitely. Destructive testing yields a value of 6.4 A at which point the electrodes begin emitting tendrils of smoke.

Figure 3g shows the SEM image of the Ag coated scaffolds, and the tip of a random cantilever microfiber is imaged in Fig. 3h. As is evident, the fibers are thoroughly covered with Ag microparticles even at the ends of the fibers and the inset to Fig. 3h shows the extent of metallization using energy selective backscattered (EsB) electron detection (scale bar 2 μm).

## Cu electroplating

Although excellent results can be achieved using the aforementioned method of Ag microparticle coating, as shown in Fig. 3a–h, it would be economically beneficial if Ag is replaced with Cu. Direct adhesion of Cu with this method however is not straightforward since Cu in a dispersion needs to be protected from oxidation until the time of coating. This demands either chemically stabilizing Cu or coating the particles with an oxygen barrier which tends to increase the complexity of the process.

Here, we implement a facile method for minimizing the use of Ag and implementing Cu without compromising on the high conductivity and quasi-transparency of the resulting Cu coated electrodes. This is done by first diluting the Ag ink with organic solvent (details in Methods) to reduce its viscosity by a full order of magnitude without changing the original Ag microparticle content. This results in a proportionally low conductivity coating when the CDT treated scaffolds are metallized (~2 mg Ag/ 2.5 cm² scaffold). The sheet resistance increases by about two orders of magnitude (~70 Ω/□) compared to what is achieved with the undiluted ink (0.5 Ω/□) (~7 mg Ag/ 2.5 cm² scaffold), as can be seen at the zero-second mark in Fig. 3i. This minimal amount of Ag acts as a seed layer for the deposition of Cu during electroplating in a CuSO₄.5H₂O bath. This facile process results in a Cu based metallization of lignocellulose scaffolds and the electrode resulting from this process is shown in Fig. 3j. Figure 3i also shows the change in sheet resistance over the electrodeposition time. A longer duration of applied power (~200 mW) results in a thicker Cu coating and the results shown in Fig. 3i indicate that a conductivity comparable to that achieved with the purely Ag-based process can be achieved within 5–7 min for a $2.5 \times 2.5\ cm^2$ scaffold. The Cu coated scaffold shown in Fig. 3j underwent electroplating for 45 s and hence still shows the underlying Ag seed layer. Figure 3k shows the bending measurements performed on the samples where Cu electroplating was performed on the Ag seed layers. The results show that the sheet resistance increases to twice the values shown in Fig. 3c (for purely Ag coated scaffolds) after 6000 bending cycles on 1 cm and 0.5 cm rollers. This indicates that, although electroplating is effective, the adhesion of the electroplated Cu requires optimization for higher mechanical reliability. This improvement in adhesion of electroplated Cu on Ag may be achieved based on previously reported techniques[28,29]. Figure 3l shows that the deposition of a Cu layer over the Ag coat reduces the transparency of the conducting scaffold by about 5%.

## Effect on pathogens

The Oligodynamic Effect refers to the phenomenon[30] that some metals display highly toxic effects on microbial cells. Metals most notably examined for their oligodynamic properties are generally heavy metals like copper[31], gold[32], silver[33], zinc[34], and their alloys (for example with magnesium)[35]. The toxic effect of metal ions on cells[36] have been correlated with multiple mechanisms such as protein denaturation, binding to mitochondrial or nuclear DNA and limiting cell multiplication[37,38], or directly causing cell membranes to fatally rupture[39,40]. Most of these mechanisms have been extremely effective against viruses[41] (including the SARS-CoV-2 virus[42]), bacteria[43], fungi[44], moulds[45], and spores[46].

Escherichia coli (E. coli) is a bacterial strain belonging to fecal coliforms, typically present in contaminated water and food, and frequently responsible for food poisoning in humans. Moreover, it also causes severe instances of urinary tract infections and gastroenteritis among humans. In turn, fecal bacteria belong to a wider group of total coliforms that are found in soil and water along with human and animal waste.

The setup used to quantify the efficacy of the metallized lignocellulose scaffolds against microbiologically contaminated water comprised stacks of these placed into containers holding contaminated water whilst under agitation for 30 min (Fig. 4a). Various parameters, such as the number of stacks (each stack contained 10 scaffolds measuring 6 cm in diameter), the type of bacterial strains, and the voltage applied across the scaffolds, were varied in this setup to study the oligodynamic efficacy.

The number of Ag coated lignocellulose scaffolds played an important role in the reduction of E. coli concentration during the 30-min duration. As shown in Fig. 4b, the bacterial reduction appears directly linked with the number of Ag coated scaffolds, since where a single stack of scaffolds exhibited an almost 50% inhibition of bacterial growth, two stacks reached an almost 100% inhibition after just 20 min of exposure. This is attributed to the increased concentration of Ag cations released into the water in proportion to the increase in Ag coated stacks and has been previously reported to be a significant contributor to the oligodynamic efficacy of metals like Ag[14].

However, the inhibition of E. coli bacterial strains using a single stack of metallized lignocellulose scaffolds reached almost 100% inhibition upon the application of 1 V DC across the stack as shown in Fig. 4c. Basically, the antibacterial effect of a single stack of scaffolds was doubled when just 1 V was applied across it and the efficacy of E. coli inhibition was close to 100% after 30 min of applying 1 V and 20 min of applying 2 V to the metallized lignocellulose electrodes, demonstrating no significant differences in bacterial inhibition (Fig. 4c). These results indicate that the oligodynamic effect, when bolstered with a current flow, can eliminate gram- negative bacterial species such as E. coli, known to be resistant against not only the standard oligodynamic effect (i.e., without the applied voltage as seen in Fig. 4b) but also against many antibiotics, soaps, detergents, and natural enzymes involved in the immune response[47–49].

Among the bacterial community fecal coliforms serve a crucial role in indicating whether aquatic systems are polluted or not. Additionally, they are utilized in the biodegradation of a diverse array of pollutants within wastewater in order to mitigate their impact on the environment. However, despite efforts, the treatment of industrial and domestic wastewater is not entirely efficient, leading to the release of pollutants and bacterial loads into

**Article**

**Fig. 4 | Oligodynamic inhibition of bacterial contaminants. a** Basic illustration of the measurement setup used to measure the oligodynamic effect on contaminated water. **b** E. coli bacterium elimination over a period of 30 min with no voltage applied to the metallized scaffolds (only one stack of metallized scaffolds was used for these experiments) (**c**) upon the application of 1 Volt (red) and 2 Volts (black) (**d**) Inhibition of fecal coliforms (FC), total coliforms (TC) and E. coli without any applied voltage (**e**) Inhibition of fecal coliforms (FC), total coliforms (TC) and E. coli with 1 V applied to the metallized lignocellulose scaffolds. The results represent the average values from at least two independent experiments and each experiment was repeated twice.

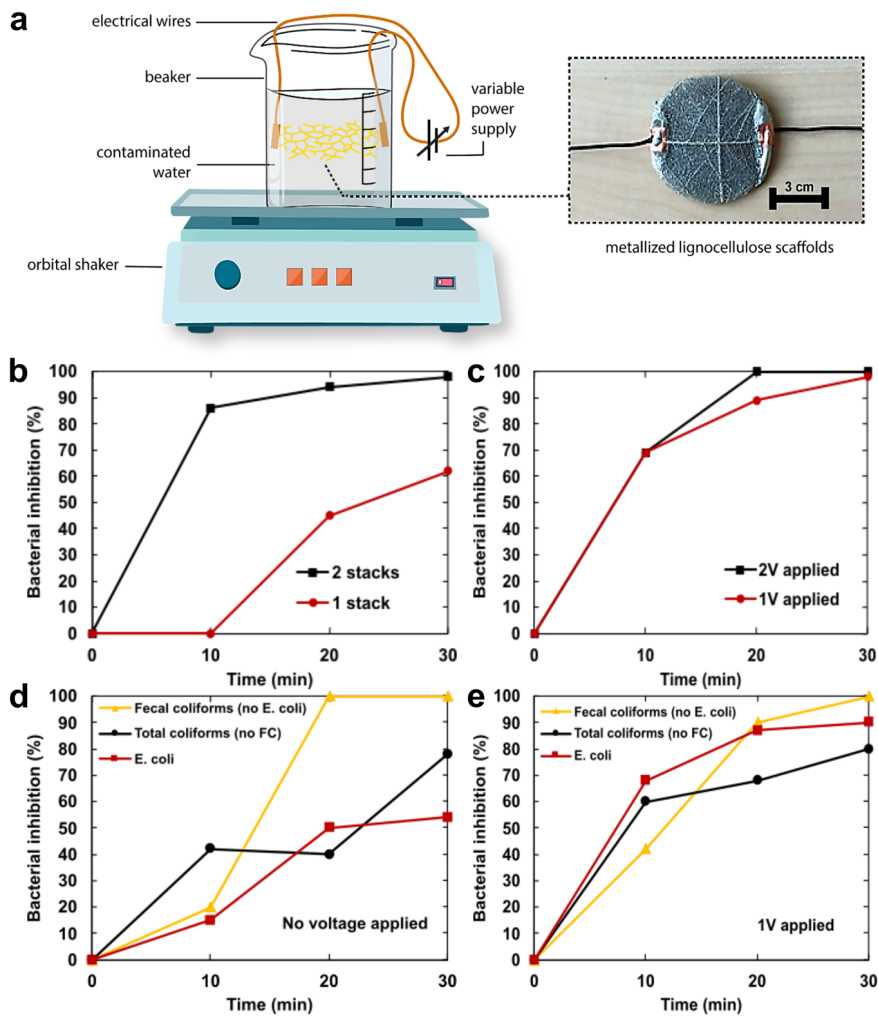

the environment. It is therefore imperative to reduce the bacterial load originating from Wastewater Treatment Plants (WWTPs) to prevent environmental contamination. Hence, the efficacy and specificity of Ag-coated lignocellulose scaffolds were assessed next against real municipal wastewater samples.

The process involved first quantifying the presence of fecal coliforms (FC), total coliforms (TC) and E. coli present in the municipal wastewater samples before implementing a single stack of metallized lignocellulose scaffolds. The bacterial attrition was subsequently measured in the absence (Fig. 4d) and presence (Fig. 4e) of an applied voltage of 1 V. In the absence of an externally applied voltage, a single stack of scaffolds could eliminate fecal coliforms with 100% efficiency within 20 min. However, the same setup simultaneously had almost no effect on the Total coliforms in 30 min. E. coli inhibition was also low with this setup and only about 50% inhibition could be achieved in 30 min (Fig. 4d). However, when a potential difference of 1 V was applied across the scaffolds, a drastic and consistent increase in pathogenic attrition rates was observed for E.coli and Total coliforms (as shown in Fig. 4e) showing an 80% inhibition of TC and about 90% inhibition of E. coli. As expected, the fecal coliform inhibition also remained high as before.

Measurements performed using Inductively Coupled Plasma Mass Spectrometry (ICP-MS) showed that 2.15 µg of Ag⁺ cations were released into 100 ml of water in 30 min irrespective of whether an external voltage was applied or not. However, the improvement in bacterial inhibition upon the application of a voltage may be caused by the electrons injected into the quasi-fractal surface of the metallized scaffolds creating oxidation sites that interfere with ion flux across bacterial cell membranes. The very rough surface of the electrodes (Fig. 3g, h) on a sub 100 µm scale may also result in an extremely high electric field density on such sharp corners, leading to a direct electrostatic breakdown of the cell membranes. Further research is required to ascertain the precise mechanisms contributing to the high efficacy of this system under an applied voltage. The results however indicate that such economically metallized micro-fractal structures have the potential to offer a sustainable means to manufacture bio-derived sieves that can decontaminate water passing through them. The high efficacy of the approach also provides sustainable alternatives for wastewater treatment along with maintaining the potabilitiy of drinking water over long periods of time, especially in famine-stricken or arid locations of the planet.

Therefore, leaf-derived quasi-fractal lignocellulose scaffolds can be metallized using a novel Ag microparticle-based methodology that eschews the use of excessive chemical treatments or nanoparticles. The properties of the resulting self-standing, flexible, quasi-transparent electrodes are found to be at par with the state-of-the-art for transparent electrodes used in thin film electronics with the demonstration of sheet resistances below 1 Ω/□ and light transmittance of 80%. It is shown that not only is it a new method to fabricate reliable leaf-based conducting quasi-fractals, but it can also be used to purify water contaminated with bacterial pathogens like E. coli and fecal coliforms. We demonstrated a high, naturally existing propensity of Ag coated lignocellulose scaffolds towards eliminating fecal coliforms with up to 100% efficacy. We also show that this antibacterial effect can be significantly widened to cover bacterial groups such as E. coli and other groups belonging to total coliforms by the application of a small electrical potential (1 V) to the Ag coated lignocellulose scaffolds. The low-cost fabrication and ease of processing makes natural lignocellulose quasi-fractals present in leaves an excellent base for further applications in the field of sustainable flexible electronics as well as water treatment.

## Data availability
The data that support the plots in this paper and other findings of this study are available from the corresponding authors upon request.

## References
1. Han, B. et al. Bio-inspired networks for optoelectronic applications. *Nat. Commun.* **5**, 5674 (2014).
2. Koivikko, A., Lampinen, V., Yiannacou, K., Sharma, V. & Sariola, V. Biodegradable, flexible and transparent tactile pressure sensor based on rubber leaf skeletons. *IEEE Sens.* J. **22**, 11241–11247 (2022).
3. Elsayes, A. et al. Plant-based biodegradable capacitive tactile pressure sensor using flexible and transparent leaf skeletons as electrodes and flower petal as dielectric laye. *Adv. Sustain. Syst.* **4**, 2000056 (2020).
4. Jia, G. Biomimic vein-like transparent conducting electrodes with low sheet resistance and metal consumption. *Nano-Micro Lett.* **12** (2020).
5. Sharma, V., Koivikko, A., Yiannacou, K., Lahtonen, K. & Sariola, V. Flexible biodegradable transparent heaters based on fractal-like leaf skeletons. *npj Flex Electron.* **4** (2020).
6. Golestanirad, L., Analysis of fractal electrodes for efficient neural stimulation, *Front. Neuroeng.* **6** (2013).
7. Watterson, W. J., Montgomery, R. D. & Taylor, R. P. Fractal electrodes as a generic interface for stimulating neurons. *Sci. Rep.* **7** (2017).
8. Tian, L. et al. Large-area MRI-compatible epidermal electronic interfaces for prosthetic control and cognitive monitoring. *Nat. Biomed. Eng.* **3**, 194–205 (2019).
9. Browning, L. A. et al. Investigation of fractal carbon nanotube networks for biophilic neural sensing applications. *Nanomaterials* **11**, 636 (2021).
10. James, S & Contractor, R. Study on nature-inspired fractal design-based flexible counter electrodes for dye-sensitized solar cells fabricated using additive manufacturing. *Sci Rep.* **8** (2018).
11. Wang, X., Wang, R., Zhai, H., Shi, L. & Sun, J. 'Leaf vein' inspired structural design of Cu nanowire electrodes for the optimization of organic solar cells. *J. Mater. Chem. C* **6**, 5738–5745 (2018).
12. Thekkekara, L. V. & Gu, M. Bioinspired fractal electrodes for solar energy storages. *Sci. Rep.* **7**, 45585 (2017).
13. Antsiferova, A. A., Kashkarov, P. K. & Koval'chuk, M. V. Effect of different forms of silver on biological objects. *Nanotechnol. Russ.* **17**, 155–164 (2022).
14. Yin, I. X. et al. The Antibacterial Mechanism of Silver Nanoparticles and Its Application in Dentistry. *Int. J. Nanomed.* **15**, 2555–2562 (2022).
15. Cao, C. et al. Long-term effects of environmentally relevant concentration of Ag nanoparticles on the pollutant removal and spatial distribution of silver in constructed wetlands with Cyperus alternifolius and Arundo donax. *Environ. Pollut.* **252**, 931–940 (2019).
16. Hund-Rinke, K. et al. Prioritising nano- and microparticles: identification of physicochemical properties relevant to toxicity to Raphidocelis subcapitata and Daphnia magna. *Environ. Sci. Eur.* **34**, 116 (2022).
17. Kruszewski, M. et al. Chapter five - toxicity of silver nanomaterials in higher eukaryotes. *Adv. Mol. Toxicol.* **5**, 179–218 (2011). Elsevier.
18. Reshmy, R. et al. Bioplastic production from renewable lignocellulosic feedstocks: a review. *Rev. Environ. Sci. Biotechnol.* **20**, 167–187 (2021).
19. Benhadi, S. et al. Corona discharge treatment route for the grafting of modified β-cyclodextrin molecules onto cellulose. *J. Incl. Phenom. Macrocycl. Chem.* **70**, 143–152 (2011).
20. Akanda, M. R., Osman, A. M., Nazal, M. K. & Aziz, M. A. Review— recent advancements in the utilization of Indium Tin Oxide (ITO) in electroanalysis without surface modification. *J. Electrochem. Soc.* **167**, 037534 (2020).
21. Hu, Z., Wang, Z., An, Q. & Zhang, F. Semitransparent polymer solar cells with 12.37% efficiency and 18.6% average visible transmittance. *Sci. Bull.* **65**, 131–137 (2020).
22. Cisneros-Contreras, I. R., Muñoz-Rosas, A. L. & Rodr, A. Resolution improvement in Haacke's figure of merit for transparent conductive films. *Results Phys.* **15**, 102695 (2019).
23. Haacke, G. New figure of merit for transparent conductors. *J. Appl. Phys.* **47**, 4086–4089 (1976).
24. Fikry, M., Mohie, M., Gamal, M. A., Ibrahim, A. & Genidy, G. Superior control for physical properties of sputter deposited ITO thin-films proper for some transparent solar applications. *Opt. Quant. Electron* **53**, 122 (2021).
25. Valle, M. et al. Design, growth, and characterization of crystalline copper oxide p-type transparent semiconductive thin films with figures of merit suitable for their incorporation into translucent devices. *Cryst. Growth Des.* **22**, 2168–2180 (2022).
26. Huang, J., Ho, Y. & Wu, Y. Effect of substrate temperature on the performance of ultrasonically sprayed PEDOT:PSS–AgNWs transparent conductive films for electrochromic devices. *J. Mater. Sci. Mater. Electron* **33**, 19490–19500 (2022).
27. Anand, A., Islam, M. M., Meitzner, R., Schubert, U. S. & Hoppe, H. Introduction of a novel figure of merit for the assessment of transparent conductive electrodes in photovoltaics: exact and approximate form. *Adv. Energy Mater.* **11** (2021).
28. Miller, A., Yu, L., Blickensderfer, J. & Akolkar, R. Electrochemical copper metallization of glass substrates mediated by solution-phase deposition of adhesion-promoting layers. *J. Electrochem. Soc.* **162**, D630 (2015).
29. Ran, W., Liu, F., Li, H. & Liu, S. Improving bonding strength between Ni/Cu/Ag coatings and MgTiO3 ceramic resonator by alumina thin-film grown by atomic layer deposition. *Ceram. Int.* **49**, 23788–23795 (2023).
30. Nägeli, C. Über oligodynamische Erscheinungen in lebenden Zellen, in *Denkschriften, Neue, d. allg. Schweiz. Gesellschaft f. d. gesammten Naturwissenschaften*, Zurich, Zürcher & Furrer, p. 51. (1893).
31. Ermini, M. L. & Voliani, V. Antimicrobial nano-agents: the copper age. *ACS Nano* **15**, 6008–6029 (2021).
32. Verma, D. K., Recent development and importance of nanoparticles in disinfection and pathogen control. in *Nanomaterials for Environmental and Agricultural Sectors. Smart Nanomaterials Technology* p. 83–106 (Springer, 2023).
33. Mittapally, S., Taranum, R. & Parveen, S. Metal ions as antibacterial agents. *JDDT* **8**, 411–419 (2018).
34. Son, J., Oh, J. K., Cho, D. H., Akbulut, M. & Teizer, W. Bacterial inactivation characteristics of magnesium–calcium–zinc alloys for bone implants. *MRS Commun.* **10**, 609–612 (2020).
35. Shao, Y. et al. Advance in antibacterial magnesium alloys and surface coatings on magnesium alloys: a review. *Acta Metall. Sin. (Engl. Lett.)* **33**, 615–629 (2020).
36. Rtimi, S., Nadtochenko, V., Khmel, I., Bensimon, M. & Kiwi, J. First unambiguous evidence for distinct ionic and surface-contact effects during photocatalytic bacterial inactivation on Cu–Ag films: Kinetics, mechanism and energetics. *Mater. Today Chem.* **6**, 62–74 (2017).
37. Rtimi, S. et al. ZrNO–Ag co-sputtered surfaces leading to E. coli inactivation under actinic light: Evidence for the oligodynamic effect. *Appl. Catal. B Environ.* **138–139**, 113–121 (2013).
38. Yang, K., Ren, L. & Yu, Y. Design and development of antibacterial metal implants. in *Revolutions in Product Design for Healthcare* p. 163–175 (Springer, 2022).

39. Salleh, A. et al. The potential of silver nanoparticles for antiviral and antibacterial applications: a mechanism of action. *Nanomaterials* **10**, 1566 (2020).

40. Obafunmi, T., Ocheme, J. & Gajere, B. Oligodynamic effect of preciousmetals on skin bacteria. *Fudma J. Sci.* **4**, 601–608 (2020).

41. Boivin, L. & Harvey, P. D. Virus management using metal–organic framework-based technologies. *ACS Appl. Mater. Interfaces* **15**, 13844–13859 (2023).

42. van Doremalen, N. et al. Aerosol and surface stability of SARS-CoV-2 as compared with SARS-CoV-1. *N. Engl. J. Med* **382**, 1564–1567 (2020).

43. Valappil, S. P. et al. Effect of silver content on the structure and antibacterial activity of silver-doped phosphate-based glasses. *Antimicrob. Agents Chemother.* **51** (2007).

44. Mahdavi-Yekta, M. et al. Silver nanoparticles and quinoa peptide enriched nanocomposite films for the detoxification of aflatoxins in pistachio. *Int. J. Environ. Anal. Chem*, 1–14. https://doi.org/10.1080/03067319.2022.2118592 (2022).

45. Evans, P. D., Matsunaga, H., Preston, A. F. & Kewish, C. M. Wood protection for carbon sequestration — a review of existing approaches and future directions. *Curr. Forestry Rep.* **8**, 181–198 (2022).

46. Horn, H. & Niemeyer, B. Aerosol disinfection of bacterial spores by peracetic acid on antibacterial surfaces and other technical materials. *Am. J. Infect. Control* **48**, 1200–1203 (2020).

47. Rakhalaru, P. et al. Prevalence and Antimicrobial Resistance Profile of Diarrheagenic Escherichia coli from Fomites in Rural Households in South Africa. *Antibiotics (Basel)* **12**, 1345 (2023).

48. Omorodion Nnenna, J. P. & Omoukaro, A. J. Antimicrobial activity of anticeptic, herbal and beauty soaps on clinical isolates (Escherichia coli, Staphylococcus aureus, Pseudomonas aeruginosa and Klebsiella pneumonia). *Int. J. Chem. Biol. Sci.* **4**, 20–28 (2022).

49. Aguilar-Salazar, A. et al. Prevalence of ESKAPE bacteria in surface water and wastewater sources: multidrug resistance and molecular characterization, an updated review. *Water* **15**, 3200 (2023).

## Acknowledgements

The authors thank the Dresden Center for Nanoanalysis for granting access to the SEM facility.

## Author contributions

H.K. and K.L. supervised the research and contributed to the editing of the final manuscript. R.N. conceived and designed the study. M.N.-L., V.R.B. and C.I. established the oligodynamic testing of the electrodes. J.W. performed the transmittance measurements and provided feedback during the writing of the manuscript. T.A. performed the SEM measurements. All the authors read and revised the manuscript.

## Funding

Open Access funding enabled and organized by Projekt DEAL.

## Competing interests

The authors declare no competing interests.

## Additional information

**Correspondence** and requests for materials should be addressed to Rakesh Rajendran Nair or Karl Leo.

**Reprints and permissions information** is available at http://www.nature.com/reprints

**Publisher's note** Springer Nature remains neutral with regard to jurisdictional claims in published maps and institutional affiliations.